\documentclass[twocolumn]{aastex62}
\usepackage{amsmath}

\shorttitle{Parametrizing the Reionization History}
\shortauthors{Trac}

\begin{document}

\title{Parametrizing the Reionization History with the Redshift Midpoint, Duration, and Asymmetry}

%\correspondingauthor{Hy Trac}
\email{hytrac@andrew.cmu.edu}

\author[0000-0001-6778-3861]{Hy Trac}
\affil{McWilliams Center for Cosmology, Department of Physics, Carnegie Mellon University, Pittsburgh, PA 15213}

\begin{abstract}
A new parametrization of the reionization history is presented to facilitate robust comparisons between different observations and with theory. The evolution of the ionization fraction with redshift can be effectively captured by specifying the midpoint, duration, and asymmetry parameters. Lagrange interpolating functions are then used to construct analytical curves that exactly fit corresponding ionization points. The shape parametrizations are excellent matches to theoretical results from radiation-hydrodynamic simulations. The comparative differences for reionization observables are: ionization fraction $|\Delta x_\text{i}| \lesssim 0.03$, 21cm brightness temperature $|\Delta T_\text{b}| \lesssim 0.7\, \text{mK}$, Thomson optical depth $|\Delta \tau| \lesssim 0.001$, and patchy kinetic Sunyaev-Zel'dovich angular power $|\Delta D_\ell | \lesssim 0.1\, \mu\text{K}^2$. This accurate and flexible approach will allow parameter-space studies and self-consistent constraints on the reionization history from 21cm, CMB, and high-redshift galaxies and quasars.
\end{abstract}

\keywords{cosmology: theory -- dark ages, reionization, first stars -- galaxies: high-redshift -- methods: analytical -- numerical}

\section{Introduction} \label{sec:intro}

The reionization of hydrogen by the first stars, galaxies, and quasars is a milestone event in the first billion years. Ionizing radiation from luminous sources convert the cold and neutral gas into a warm and highly ionized medium \citep[][for a review]{2013fgu..book.....L}. Recent observations suggest that the epoch of reionization (EoR) was already in significant progress by redshift $z \sim 8$ and must have ended by $z \sim 6$ \citep[][and references therein]{2016A&A...596A.108P}. Upcoming observations will better constrain the reionization history, as well as the abundance and properties of the radiation sources.

The reionization history is quantified by the evolution of the ionization fraction $x_\text{i}(z)$. It is used in the calculation of EoR observables such as the 21cm brightness temperature \citep[e.g.][]{1997ApJ...475..429M, 2006PhR...433..181F}, Thomson optical depth, cosmic microwave background (CMB) temperature and polarization anisotropies \citep[e.g.][]{1987MNRAS.226..655B, 1997PhRvD..55.1822Z}, and kinetic Sunyaev-Zel'dovich (KSZ) effect \citep[e.g.][]{1970Ap&SS...7....3S, 1986ApJ...306L..51O}. Thus, it is important to establish a standard parametrization of the reionization history to facilitate robust comparisons between different observations and with theory.

The ionization fraction is often parametrized using a $\tanh$ function with two free parameters which set the redshift midpoint and width \citep{2008PhRvD..78b3002L}. However, the width parameter does not clearly define the duration of the EoR and the simple functional form does not allow for possible asymmetry. Redshift-asymmetric parameterizations using polynomials, exponentials, and power-laws have recently been proposed \citep[e.g.][]{2015A&A...580L...4D, 2016A&A...596A.108P}. Generalized logistic functions \citep{doi:10.1093/jxb/10.2.290} can have asymmetric sigmoid shapes, but the physical interpretation of some of the free parameters for reionization is not straightforward.

In this Letter, I present an accurate parametrization of the reionization history in terms of the redshift midpoint, duration, and asymmetry. Lagrange interpolating functions are used to construct analytical curves that exactly fit corresponding ionization points. The shape parametrizations are then compared against radiation-hydrodynamic simulations from the Simulations and Constructions of the Reionization of Cosmic Hydrogen (SCORCH) project \citep{2015ApJ...813...54T, 2017arXiv171204464D}. The adopted cosmological parameters are: $\Omega_{\rm m} = 0.3$, $\Omega_\Lambda = 0.7$, $\Omega_{\rm b} = 0.045$, $h = 0.7$, $\sigma_8 = 0.8$, $n_{\rm s} = 0.96$, $Y_\text{He} = 0.24$, and $T_\text{CMB} = 2.725$ K.

\clearpage
\section{Methods} \label{sec:method}

\subsection{Midpoint, Duration, and Asymmetry}

The reionization history is quantified by the ionized hydrogen fraction, which can be mass-weighted or volume-weighted. I will work with the mass-weighted version $x_\text{i,M}$ as the volume-averaged ionized hydrogen number density is given by
\begin{equation}
\bar{n}_\text{HII,V} = x_\text{i,M} \bar{n}_\text{H,V} .
\end{equation}
From here on, the mass-weighted and volume-averaged subscripts will be dropped to simplify the notation. Also, let $z_\text{x}$ denote the redshift corresponding to the ionization factor $\text{x}=100x_\text{i}$. 

In \citet{2017arXiv171204464D}, we choose the redshift midpoint as $z_{50}$ and present two practical choices for defining the duration $\Delta_\text{z}$ and asymmetry $A_\text{z}$ parameters. In the first case:
\begin{align}
\Delta_\text{z50} & \equiv z_{25} - z_{75} , \nonumber \\
A_\text{z50} & \equiv \frac{z_{25} - z_{50}}{z_{50} - z_{75}} ,
\end{align}
the redshifts correspond to quartile ionization fractions ($x_\text{i} = 0.25, 0.50, 0.75$) and $\Delta_\text{z50}$ is analogous to a full width half max. In the second case:
\begin{align}
\Delta_\text{z90} & \equiv z_{05} - z_{95} , \nonumber \\
A_\text{z90} & \equiv \frac{z_{05} - z_{50}}{z_{50} - z_{95}} ,
\end{align}
the redshifts correspond to early- and late-ionization fractions ($x_\text{i} = 0.05, 0.95$) and $\Delta_\text{z90}$ effectively quantifies the full extent of the EoR. While other definitions (e.g.~$\Delta_\text{z68}, \Delta_\text{z95}$) can be adopted, extreme choices (e.g.~$\Delta_\text{z99}$) are not recommended because the start and end of the EoR are difficult to determine precisely.

\subsection{Lagrange Interpolating Functions}

Given the midpoint, duration, and asymmetry parameters, the relevant redshifts are uniquely specified and given by
\begin{align}
z_{25} & = z_{50} + \frac{\Delta_\text{z50} A_\text{z50}}{1 + A_\text{z50}} , \nonumber \\
z_{75} & = z_{25} - \Delta_\text{z50},
\end{align}
or
\begin{align}
z_{05} & = z_{50} + \frac{\Delta_\text{z90} A_\text{z90}}{1 + A_\text{z90}} , \nonumber \\
z_{95} & = z_{05} - \Delta_\text{z90} .
\end{align}

An analytical function that exactly passes through a given set of ionization points and therefore satisfies the chosen midpoint, duration, and asymmetry parameters can be constructed using the method of Lagrange interpolation. In practice, a straightforward interpolation of $x_\text{i}$ in terms of $z$ or $1+z$ can have oscillations, but the approach works better with a simple change of variables,
\begin{align}
u & = \ln(1+z) , \nonumber \\
v & = \ln x_\text{i} .
\end{align}
For $N$ points, a polynomial $v(u)$ of degree $N-1$ can be constructed as
\begin{align}
v(u) & = \sum_{j=1}^N p_j(u) , \nonumber \\
 p_j(u) & = v_j\prod_{\substack{k = 1 \\ k \ne j}}^N \frac{u - u_k}{u_j - u_k} .
\end{align}
The ionization fraction $x_\text{i}(z) = \exp[v(u)]$ has the advantages of being continuous, differentiable, integrable, and invertible. At higher redshifts toward the start of the EoR, $x_\text{i}$ asymptotically goes to zero as required. At lower redshifts after the end of reionization, a physical maximum limit of unity should be imposed in practice.

\subsection{Radiation-Hydrodynamic Simulations}

\begin{deluxetable}{lCCCCC}[t!]
\tablewidth{\hsize}
\tablecaption{\label{tab:sims} The redshift midpoint, duration, and asymmetry parameters for three RadHydro simulations from the SCORCH project.}
\tablehead{
\colhead{Model} & \colhead{$z_{50}$} & \colhead{$\Delta_\text{z50}$} & \colhead{$\Delta_\text{z90}$} & \colhead{$A_\text{z50}$} & \colhead{$A_\text{z90}$}
}
\startdata
Sim 0 & 7.95 & 1.87 & 4.68 & 1.63 & 2.90 \\
Sim 1 & 7.91 & 2.27 & 5.45 & 1.59 & 2.69 \\
Sim 2 & 7.83 & 2.89 & 6.54 & 1.49 & 2.33
\enddata
\end{deluxetable}

To test the accuracy of the analytical parametrizations, I compare them against simulation results from the SCORCH project. In \citet{2017arXiv171204464D}, we present three reionization simulations with the same galaxy luminosity functions, but with different radiation escape fraction $f_{\rm esc}(z)$ models. The simulations are designed to have fixed Thomson optical depth $\tau \approx 0.06$, consistent with recent CMB observations \citep[e.g.][]{2016A&A...596A.108P, 2016A&A...596A.107P}.

The simulations are run with the RadHydro code, which combines N-body and hydrodynamic algorithms \citep{2004NewA....9..443T} with an adaptive raytracing algorithm \citep{2007ApJ...671....1T} to directly and simultaneously solve collisionless dark matter dynamics, collisional gas dynamics, and radiative transfer of ionizing photons. Each RadHydro simulation has $2048^3$ dark matter particles, $2048^3$ gas cells, and up to 12 billion adaptive rays in a comoving box of side length $50\ h^{-1}$Mpc. 

Table \ref{tab:sims} lists the midpoint, duration, and asymmetry parameters for the three SCORCH simulations. The index in the model name reflects the power-law slope in the evolution of the radiation escape fraction with $1+z$. Sim 0 has constant $f_{\rm esc}$ and reionization starts latest, but ends earliest out of the three models. Sim 1 has $f_{\rm esc}(z)$ varying linearly and is an intermediate model. Sim 2 has $f_{\rm esc}(z)$ varying quadratically and reionization starts earliest, but ends latest.

\section{Results}

\subsection{Ionization Fraction}

\begin{figure}[t!]
\center
\includegraphics[width=\hsize]{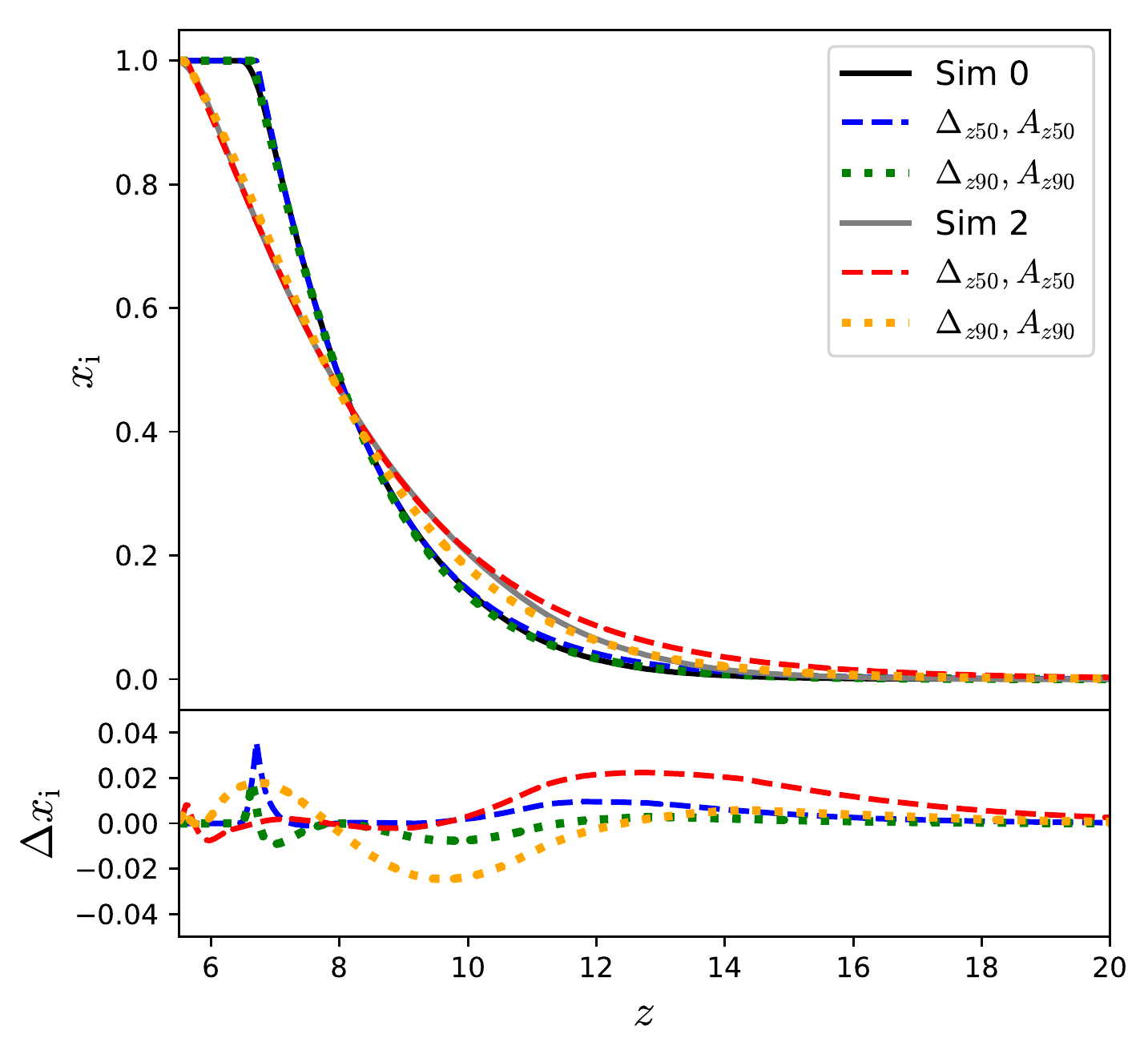}
\caption{{\bf Top:} The evolution of the mass-weighted ionization fraction with redshift. The analytical parametrizations accurately capture the redshift-asymmetric form of the simulation curves. {\bf Bottom:} The typical differences in ionization fractions are $|\Delta x_\text{i}| \lesssim 0.02$, while the maximum differences of $|\Delta x_\text{i}| \lesssim 0.03$ are found near the start and end of the EoR, which are uncertain in the simulations.}
\label{fig:ionization}
\end{figure}

Figure \ref{fig:ionization} shows the evolution of the ionization fraction for the redshift range $5.5 < z < 20$. Only Sim 0 and Sim 2 are shown for clarity as Sim 1 gives intermediate results. The analytical parametrizations are excellent matches to the simulation results and the typical differences are only $|\Delta x_\text{i}| \lesssim 0.02$. The maximum differences of $|\Delta x_\text{i}| \lesssim 0.03$ are found near the start and end of the EoR, which are also not accurately captured in reionization simulations and semi-analytical models.

The shape parametrizations using $\Delta_\text{z50}$ and $A_\text{z50}$ produce more accurate results near the midpoint, while those with $\Delta_\text{z90}$ and $A_\text{z90}$ produce smaller differences near the start and end of the EoR as expected. More accurate fits to simulation results can be obtained by combining both cases and using five rather than three ionization points. However, for parameter-space studies and constraining reionization histories from different observations, it is preferable to use a smaller number of free parameters to reduce degeneracies.

\subsection{21cm Brightness Temperature}

The global 21cm brightness temperature \citep[e.g.][]{1997ApJ...475..429M} in units of mK is given by
\begin{equation}
\delta T_\text{b} \approx 28 x_\text{HI}\left(1-\frac{T_\gamma}{T_\text{s}}\right)\left(\frac{\Omega_\text{b}h^2}{0.022}\right)\left[\left(\frac{0.15}{\Omega_\text{m}h^2}\frac{1+z}{10}\right)\right]^{1/2} ,
\end{equation}
where $x_\text{HI} = 1 - x_\text{i}$ is the neutral hydrogen fraction, $T_\gamma$ is the radiation temperature, and $T_\text{s}$ is the spin temperature. The standard approximation $T_\text{s} \gg T_\gamma$ is used, which is a valid assumption except in the early stages of reionization \citep[e.g.][]{2008ApJ...689....1S}.

\begin{figure}[t!]
\center
\includegraphics[width=\hsize]{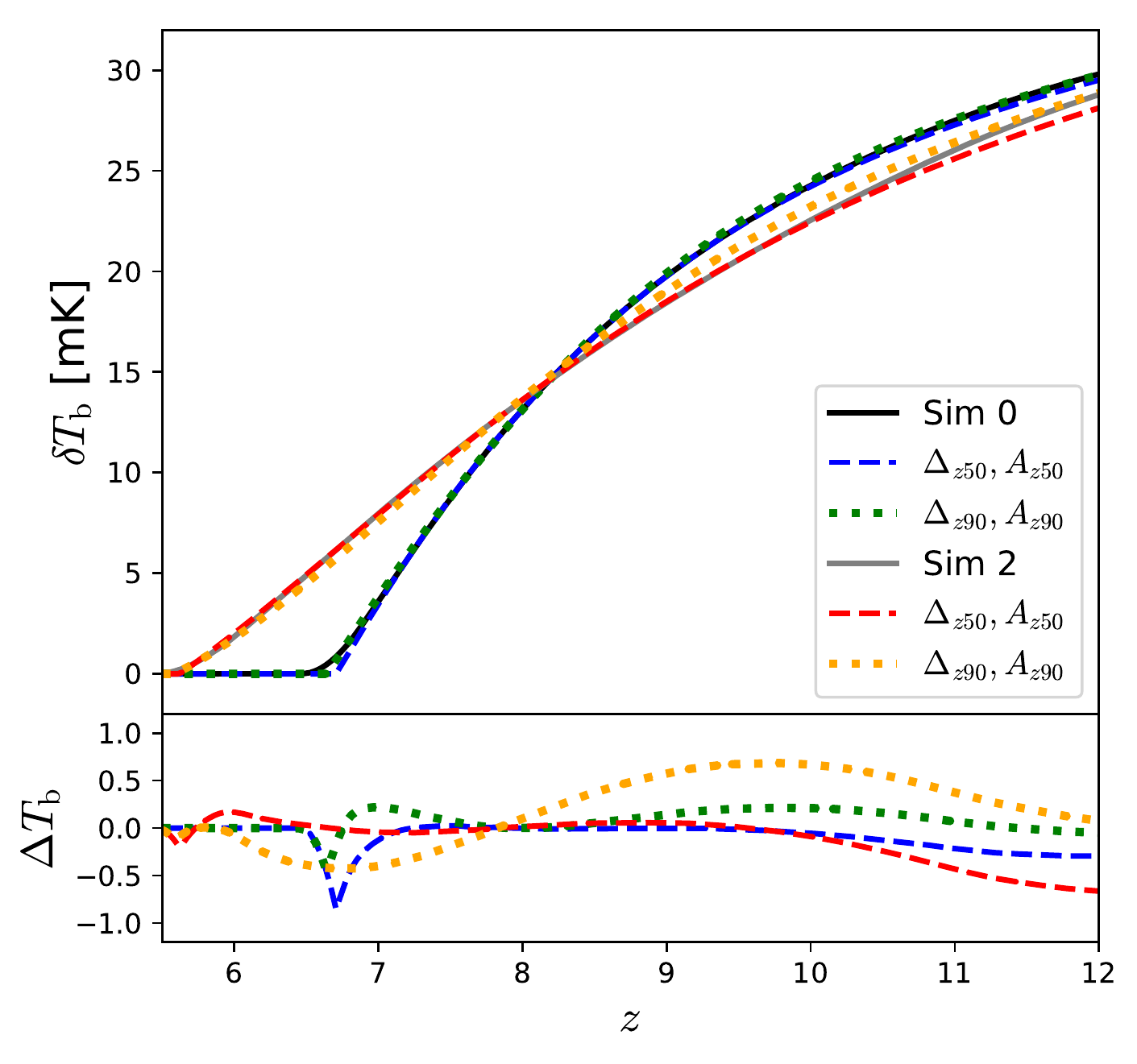}
\caption{{\bf Top:} The global 21cm brightness temperature for the redshift range excluding the early stages of reionization. {\bf Bottom:} The differences in brightness temperatures correspond to those in the ionization fractions, but have opposite signs. The typical differences are only $|\Delta T_\text{b}| \lesssim 0.7$ mK and are small compared to current observational sensitivities.}
\label{fig:21cm}
\end{figure}

Figure \ref{fig:21cm} shows the evolution of the brightness temperature for the ranges $5.5 < z < 12$ and $x_\text{HI} \lesssim 0.8$. The analytical parametrizations are excellent matches to the simulation results as expected. The differences in the brightness temperatures correspond to those in the ionization fractions, but have opposite signs. The typical differences are only $|\Delta T_\text{b}| \lesssim 0.7$ mK and are small compared to current observational sensitivities.

Global 21cm experiments such as EDGES \citep{2008ApJ...676....1B}, SCI-HI \citep{2014ApJ...782L...9V}, SARAS \citep{2017ApJ...845L..12S}, and PRIZM (Philip et al. in prep) that observe up to a frequency of 200 MHz will probe the reionization of hydrogen. To model and interpret their signals, they can use this analytical parameterization and explore the parameter space to put constraints on the redshift midpoint and duration, and possibly weaker bounds on the asymmetry.

\subsection{Thomson Optical Depth}

The Thomson optical depth integrated from redshift 0 to $z$ is given as
\begin{equation}
\tau(z) = \sigma_\text{T}\int_0^z \bar{n}_\text{e}(z) \left|\frac{cdt}{dz}\right| dz ,
\label{eqn:tau}
\end{equation}
where the volume-averaged free electron number density,
\begin{equation}
\bar{n}_\text{e} = x_\text{HII}\bar{n}_\text{H} + x_\text{HeII}\bar{n}_\text{He} + 2x_\text{HeIII}\bar{n}_\text{He}
\label{eqn:ne}
\end{equation}
is related to the mean number densities ($\bar{n}_\text{H}, \bar{n}_\text{He}$) and mass-weighted ionization fractions ($x_\text{HII}, x_\text{HeII}, x_\text{HeIII}$) for hydrogen and helium. HI and HeI are jointly ionized during the EoR \citep[e.g.][]{2007ApJ...671....1T}. While HeII reionization is also extended \citep[e.g.][]{2009ApJ...694..842M, 2017ApJ...841...87L}, the simple approximation of an instantaneous transition at $z \approx 3$ is sufficiently accurate for calculating the optical depth.

\begin{figure}[t]
\center
\includegraphics[width=\hsize]{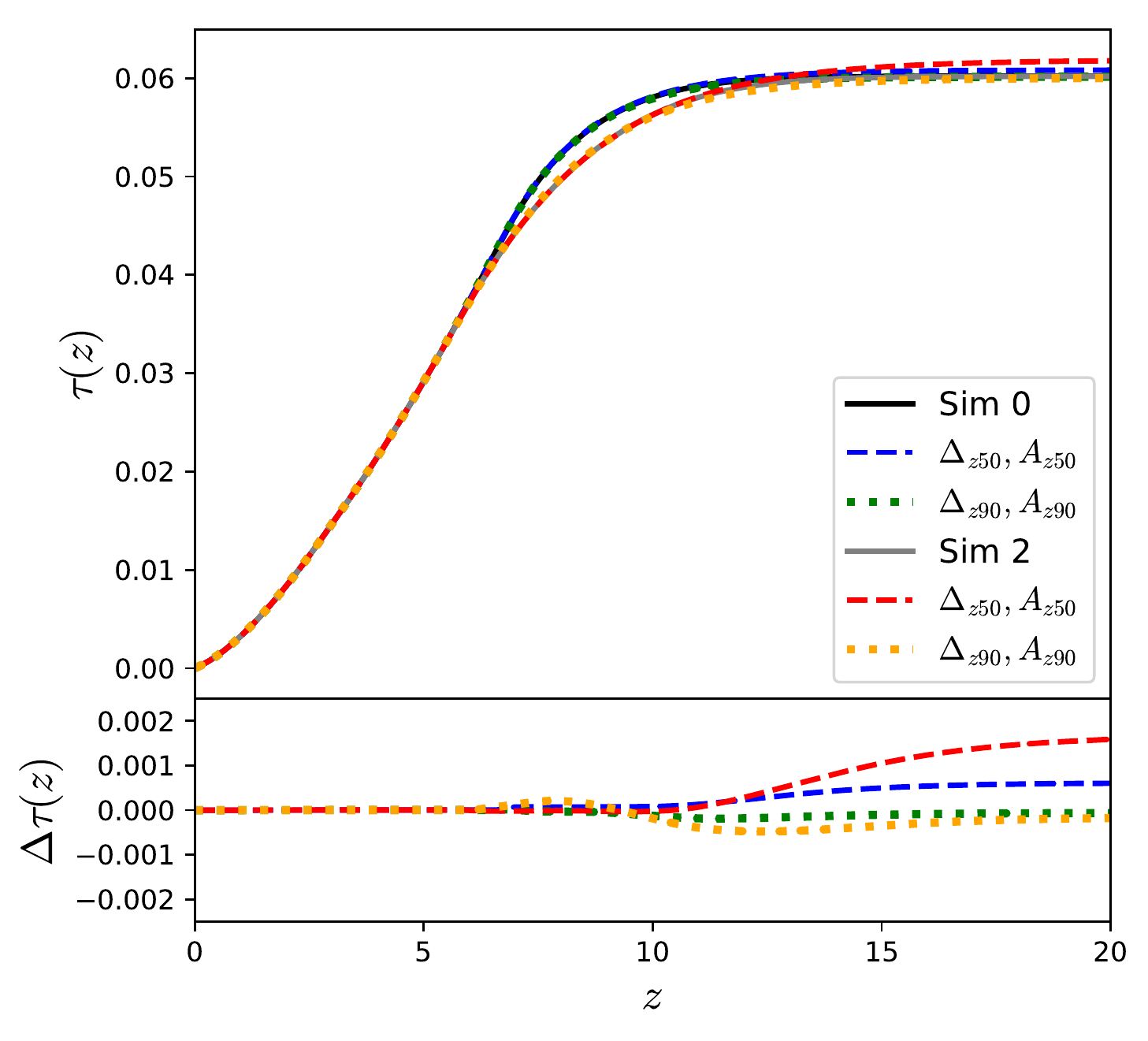}
\caption{{\bf Top:} The Thomson optical depth integrated up to redshift $z$. The SCORCH sims are designed to have fixed Thomson optical depth $\tau \approx 0.06$, consistent with recent CMB observations. {\bf Bottom:} The differences in optical depths are smaller for parametrizations using $\Delta_\text{z90}$ and $A_\text{z90}$, which more effectively quantify the full extent of the EoR.}
\label{fig:tau}
\end{figure}

Figure \ref{fig:tau} shows that the analytical parametrizations accurately reproduce the integrated optical depth $\tau$ from the simulations with typical differences of $|\Delta \tau| \lesssim 0.001\ (\lesssim 2\%)$. The shape parametrizations using $\Delta_\text{z90}$ and $A_\text{z90}$ produce very small differences of $|\Delta \tau| \lesssim 2\times10^{-4}$ because the differences in the ionization fraction $\Delta x_\text{i}$ have both positive and negative values that average to nearly zero over the EoR redshift range. For integrated statistics that are linear in $x_\text{i}$, I recommend using the parameterizations $\Delta_\text{z90}$ and $A_\text{z90}$.

Planck will soon provide an update on their current constraint of $\tau = 0.058 \pm 0.012$ \citep{2016A&A...596A.108P} from measurements of the CMB temperature and polarization angular power spectra \citep{2016A&A...596A.107P}. Since the location and amplitude of the reionization bump is precisely set by the redshift midpoint and optical depth, using an analytical parametrization that exactly matches a given $z_{50}$ and accurately produces a desired $\tau$ is highly advantageous. The current tanh function in CAMB \citep{2008PhRvD..78b3002L} can be replaced with this more accurate and flexible parametrization.

\newpage
\subsection{Patchy KSZ Effect}

The KSZ temperature distortion \citep{1970Ap&SS...7....3S} integrated along the direction ${\bf \hat{n}}$ is given by
\begin{equation}
\frac{\Delta T}{T}({\bf \hat{n}}) = -\frac{\sigma_\text{T}}{c}\int n_\text{e}({\bf v \cdot \hat{n}})e^{-\tau} \left|\frac{cdt}{dz}\right| dz ,
\end{equation}
where the electron number density $n_\text{e}$, peculiar velocity ${\bf v}$, and optical depth $\tau$ are all dependent on ${\bf \hat{n}}$ and $z$. 
In \citet{2013ApJ...776...83B}, we choose to integrate over the redshift range $5.5 \leq z \leq 20$ for the patchy KSZ component since some models can have late end to reionization, like in Sim 2 here.

To quantify the impact of small differences in $x_\text{i}(z)$ on the patchy KSZ effect, I use a new and fast semi-numerical method of modeling reionization on large scales. In Holst et al.~(in prep), we develop a novel approach that uses abundance matching to exactly satisfy a given $x_\text{i}(z)$. Density and velocity fields are constructed using 2nd-order Lagrangian perturbation theory \citep{1998MNRAS.299.1097S} with $2048^3$ particles in a periodic comoving box of side length $1\ h^{-1}$Gpc. Full-sky HEALPix \citep{2005ApJ...622..759G} maps with $N_\text{side} = 4096$ are then constructed by ray tracing through the simulated light cones.

\begin{figure}[t]
\center
\includegraphics[width=\hsize]{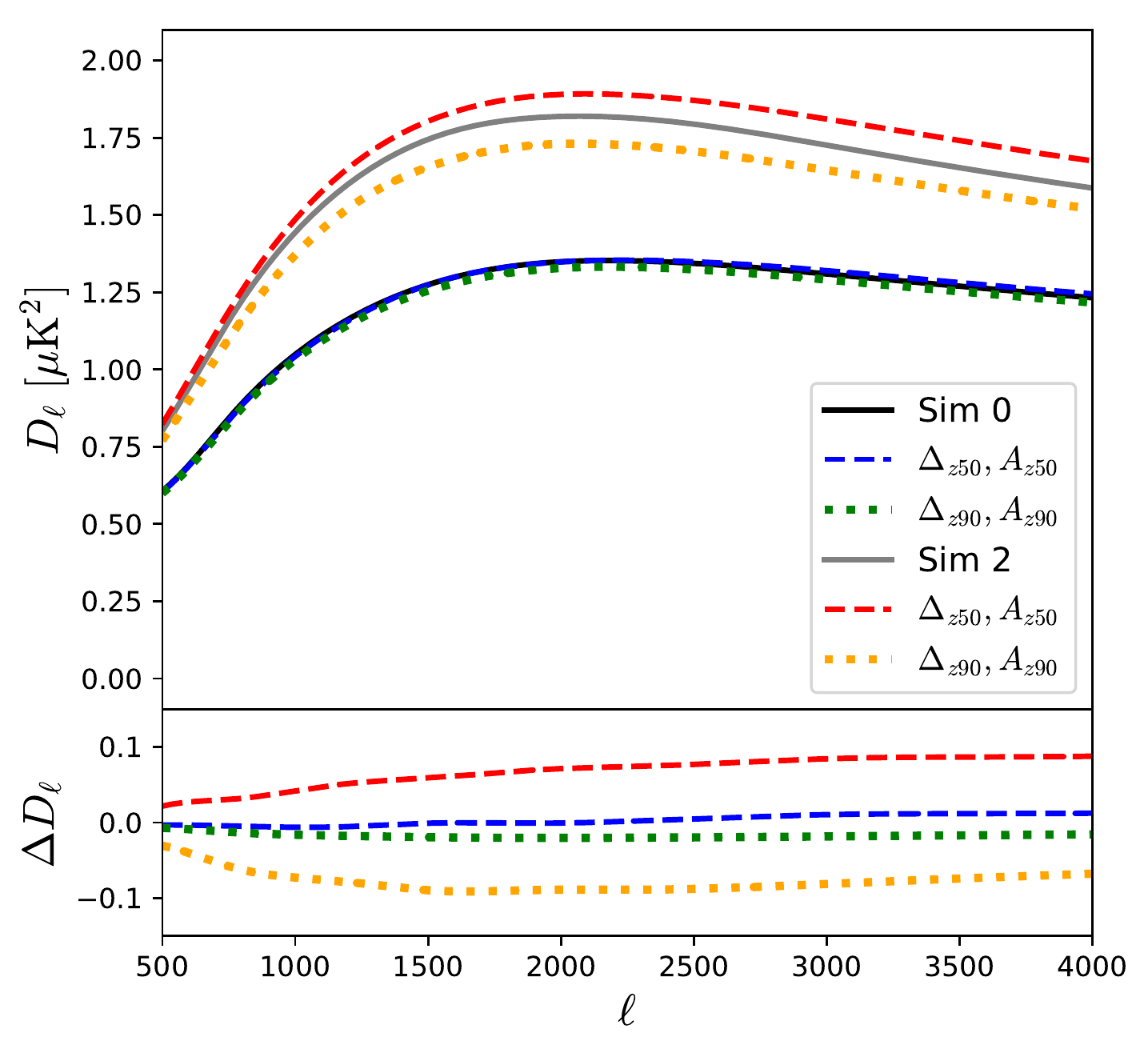}
\caption{{\bf Top:} The patchy KSZ angular power spectrum for temperature fluctuations integrated over the redshift range $5.5 < z < 20$. The overall amplitude increases for longer duration. {\bf Bottom:} The differences in angular power have similar absolute values for both parametrizations, but there are larger differences for cases with larger $|\Delta x_\text{i}|$ and longer durations.}
\label{fig:ksz}
\end{figure}

Figure \ref{fig:ksz} shows the angular power spectrum $D_\ell \equiv \ell(\ell + 1)C_\ell/(2\pi)$ in units of $\mu\text{K}^2$. The overall amplitude is expected to increase with both the redshift midpoint and duration \citep[e.g.][]{2012ApJ...756...65Z, 2013ApJ...776...83B}. The shape parametrizations using $\Delta_\text{z50}$ and $A_\text{z50}$ overpredict, while the those using $\Delta_\text{z90}$ and $A_\text{z90}$ underpredict compared to the simulations. These trends correspond to the differences in the ionization fractions at higher redshifts $z > z_{50}$. There are also larger absolute differences for Sim 2 than Sim 0 because of the larger differences in $|\Delta x_\text{i}|$ and the longer duration. The maximum differences of $\Delta D_\ell \lesssim 0.1\ \mu\text{K}^2$ ($\lesssim 5\%$) are expected to be atypical since Sim 2 has a rather long duration and late end to reionization at $z \approx 5.5$. In upcoming work, I will explore the dependence of the patchy KSZ effect on the midpoint, duration, and asymmetry parameters. 

\citet{2016A&A...596A.108P} combined their $\tau$ constraints with South Pole Telescope measurements of the KSZ angular power at $\ell = 3000$ \citep{2015ApJ...799..177G}, along with our KSZ theoretical models \citep{2013ApJ...776...83B} to infer a duration $\Delta_\text{zCMB} \equiv z_{10} - z_{99} < 2.9$ (95\% confidence interval). In \citet{2017arXiv171204464D}, we find that the upper limit on the duration is in tension with our radiation-hydrodynamic simulations, all of which have longer durations. The current discrepancy most likely is due to assumptions made in the analyses, models, and simulations. Other contributing factors could be inconsistencies in parametrizing the reionization history and ambiguity in mass-weighted and volume-weighted ionization fractions.

\section{Conclusions}

I present an accurate parametrization of the reionization history in terms of the redshift midpoint, duration, and asymmetry. Lagrange interpolating functions are used to construct analytical curves that exactly fit corresponding ionization points. I recommend using the shape parameters $\Delta_\text{z90}$ and $A_\text{z90}$ and caution against extreme choices (e.g.~$\Delta_\text{z99}$) since the start and end of the EoR are difficult to determine precisely. More accurate fits to simulation results can be obtained by using more ionization points, but a smaller number of free parameters is preferable for fitting observations. This accurate and flexible approach will allow parameter-space studies and self-consistent constraints on the reionization history from 21cm, CMB, and high-redshift galaxies and quasars.

\acknowledgments

I thank Nick Gnedin for the original motivation for this work. Thanks also go to Marcelo Alvarez, Nick Battaglia, Aristide Doussot, Ian Holst, and Sasha Kourov for helpful discussions. This work is supported by STScI grant HST-AR-15013.002-A.

%\bibliographystyle{aasjournal}
%\bibliography{reionization}

\end{document}